\documentclass{article}

\usepackage[english]{babel}

\usepackage[letterpaper,top=2cm,bottom=2cm,left=3cm,right=3cm,marginparwidth=1.75cm]{geometry}

\usepackage{amsmath}
\usepackage{graphicx}
\usepackage[colorlinks=true, allcolors=blue]{hyperref}

\title{Random Test Generation of Application Programming Interfaces}
\author{Eitan Farchi, Krithika Prakash, Vitali Sokhin}

\begin{document}
\maketitle

\begin{abstract}
Cloud high quality API (Application Programming Interface) testing is essential for supporting the API economy. \textit{Autotest Assist} is a random test generator that addresses this need.  It reads the API specification and deduces a model used in the test generation.  This paper describes \textit{Autotest Assist}.  It also address the topic of API specification pitfalls which \textit{Autotest Assist} may reveal when reading the specification.  A best practice is to add an appropriate test to the regression once a problem is revealed and solved. How to do that in the context of Autotest Assist's random test generation is covered.  
%We also discuss appropriate coverage models for API testing and how they can guide the random test generation. 
\end{abstract}

\section{Introduction}

API economy is driving the digital transformation of business applications across the hybrid cloud and edge environments. In such applications, micro-services are typically defined and expose a set of API interfaces.  Each micro service serves a specific business need.  By composing the micro services from multiple vendors a greater business value is obtained.   For example, a book store may have an inventory micro-service the buyer can search and a shopping cart micro-service that keeps an inventory of the books the buyer would like to buy.  In addition, a credit validation micro-service is leveraged from another vendor and a book shipping micro-service by yet another vendor.  Together, the micro-services allow the buyer to search the bookstore, decide on the books she would like to buy, and then buy them.  The micro-services may be placed in hybrid Cloud environments across edge locations and multiple Cloud environments and are potentially accessed by the buyer from an edge device.  In such environments, in addition to the functional requirements, communication failures, order of message passing between the micro services, placement of the micro services and failure to react by a micro service (sometimes referred to as circuit breaking) may result in quality issues.  Thus, in the API economy testing requirements should be addressed to create a high quality solution.  

An API call results in obtaining some resources that are then used to call another API that belongs to the same or some other micro service thus obtaining a business goal.  In addition, an API interface constitute of several input and output parameters.  Thus, the possible sequences of API calls with different parameters values create a huge space of possible user scenarios of the APIs.  As a result, it is not practical to attempt to exhaustively execute all of the possible user scenarios.  Traditional approaches to testing include manually defining use cases that include calls to a sequence of APIs and manually choosing parameters values using boundary conditions selection.  This process is sometimes called directed testing and is labor intensive.  To ease some of the effort involved in the testing of APIs we advocate the use of random test generation instead of the manual design of test scenarios.

Under the random test generation paradigm we randomly choose an API to execute, f(), and then randomly choose legal input parameters $p_1, \ldots, p_k$, for f(). Next, f() is run and the output and side effects that occurred in the system are observed.   This process is repeated thus generating a random sequence of API calls.  The following challenges need to be tackled when following the random test generation paradigm. 
\begin{enumerate}
    \item The input $p_1, \ldots, p_k$ should be syntactically correct.  Typically, this requirement is easy to meet.  A much stronger requirement that is harder to meet is that the inputs $p_1, \ldots, p_k$ should adhere to the preconditions of the API f() if we intend that the call to f() succeeds.   For example, we may obtain a book from the inventory using one API, f(), and then attempt to buy it using another API, g().  When calling the API g() we need to hold a valid reference to a book, which is a pre-condition calls to g() need to be meet.
    \item Be able to determine if the call to the API behaved as expected. 
    \item If a problem occurred as a result of the random generation process, the system needs to support its debugging. 
    \item Given that we are also creating directed tests and have a test suite composed of directed tests that we use to regress the system, how do we integrate a new test obtained from the random test generation to the directed regression suite? This question especially applies when the random test generation revealed a problem and once the problem is debugged we would like to enhance the regression with an appropriate test.
    \item How do we determine the coverage obtained by the random generation process?  Can we trust the random generation process to regress the system or should we still develop a directed test suite?
\end{enumerate}

To address challenge one and two above we read the API specification.  Implicitly that requires that the API specification defines its pre and post conditions.  Indeed one of the results of our approach which we elaborate on below is a set of guidelines and best practices to the definition of the API specification.

The third and forth challenges above are related.  We keep a database (DB) of the scenarios that the random test generation created.  This DB is queried to assist in the debugging and creation of an appropriate directed test that is then added to the directed regression suite.

The last point has long term consequences on maturity of your random test generation approach. If you can trust your coverage criterion, e.g., calling all pairs of dependent APIs through the output of a resource by one API used by the other API, you may regress the system by calling the random test generation until that coverage criterion is met.  We discuss that high level of mature usage of random test generation below.

The rest of this paper is organized as follows.  We first discuss API specification and design and how it impacts effective API testing. Then we introduce random test generation.  Next we cover Autotest Assist, a specific implementation of the random test generation paradigm.  We conclude by addressing the methodological challenges mentioned above in detail.

\section{Functional random test generation}
Another way for automatically generating test cases, which is not necessarily dependant on any coverage model is using an OpenAPI-based random test generation. This method has the ability to create sequences of random API requests/responses, in a single or multiple processes concurrently.

Random test generation for functional testing of APIs has two main goals. First, it automates the usually-manual process of writing tests, which is labor-intensive, bug-prone on its own, and limited in its ability to cover the API functionality. Second, it has the ability to generate API calls and sequences which a human will not consider writing tests for, thus provide higher and unique coverage, as well as hit hard-to-detect bugs.

\subsection{Random test generation characteristics}
Over the past few years, since REST APIs became the de-facto standard for writing web applications, there have been several works on tools and technologies for automatic test generation for APIs (see \cite{Evomaster}, \cite{Restler}, \cite{Pythia}). The different tools can be characterized using several attributes. We will use these attributes to discuss tradeoffs in this area, as well as provide some examples and best practices.

\subsubsection{Testing approach}
The first aspect to consider is whether the tool uses the API source code ("white-box") or not ("black-box"). Needless to say, a white-box approach has an enormous potential for leveraging the knowledge on the code structure and the internal methods and parameters, for generating more qualitative tests. However, it is not always possible to have access to the source code (e.g. due to confidentiality). Furthermore, this technique is usually coupled with a specific code language, and is therefore limited in its usage. 

Most of the existing tools use the black-box approach, which is sub-optimal, but more generic and easy to use. Using a white-box approach can be more suitable for tools which aim at a specific code language, or at a concrete set of APIs which need to be thoroughly tested (e.g. because they are part of a mission-critical application).

\subsubsection{Test-generation technique}
There are many techniques for generating random test-cases: model-based, fuzzing, property-based, and more. There is no "right" technique, and each has its own pros and cons. Model-based generation, for example, allows for a more accurate test generation, which covers the API space better. However, it relies on a relatively elaborate specification of the API, and depends on the ability to discover connections between operations and resources in the API. Fuzzing requires less structural data, but does rely on a "good" set of existing tests which can be used as a basis for the fuzzing algorithm. Evidently, the choice of a test-generation technique is heavily influenced by the experience and knowledge of the developers of the tool, rather than an informative decision - as said, there isn't a single "best" method.

\subsubsection{Test derivation}
Some tools can derive concrete test-cases out of the randomly generated tests, to be used in regressions. This is a powerful feature, that if exists, can potentially make the manual test writing (almost) completely redundant. A random test will usually contain symbolic variables, to pass from one request/response to another, simply because the API is non-deterministic, and can return different outputs in different invocations of the same request (given the same initial state). Tools which don't have this feature, are used mainly for bug-hunting and system stressing, but can't be used for building a systematic regression suite.

\subsubsection{State tracking}
Some tools track an internal state for better generation and/or checking. This internal state is supposed to (as much as possible) accurately reflect the actual state of the system, with respect to the existing resources and connections between them. The state can be used for better generation, for example using existing resources for creating sub-resources or for deletion. Furthermore, it can be used for better checking, for example when predicting a response code of creation/deletion of an object, partially solving the infamous oracle problem in SW testing 
%(TODO - add reference).

\subsubsection{Checking method}
One of the notoriously difficult problems in software testing (and testing in general), is the oracle problem. API testing is not different, and several methods for checking exist. First, all of the tools can perform syntactic checking of responses, asserting that their format and content is valid with respect to the API definition. However, many APIs are under-speficied, so this type of checking is not that powerful. Another option is to detect all of the $500$ responses as bugs. Usually this does the trick, unless the API allows for these responses, making this check obsolete. The higher level of checking is "semantic checking", in which the tool can predict whether the response code which was returned is (one of) the correct code(s). This is much more difficult to achieve, yet highly useful in detecting bugs. 
%(TODO - add something about our existing work on that?)
 
\subsubsection{Automation level}
The different tools differ in the effort it takes to use the tool on a new API. The ideal tool is one that needs no manual configuration - it takes the API specification (and possibly other inputs, such as the source code), and automatically starts generating tests which detect bugs. However, some manual configuration is usually both needed and helpful, in order to direct the tool towards more promising directions which will find more bugs.

\section{IBM Autotest Assist Technology}
In this section we describe Autotest Assist which is a model-based biased-random test generator developed by IBM Research. In subsequent sections we will explain the testing methodology that Autotest Assist implements. Autotest Assist's initial purpose was to help IBM in developing and testing its APIs for various purposes such as IBM Public Cloud, middle-ware offerings, software tools. Later, in collaboration with the IBM Product team, Autotest Assist evolved to be an efficient testing tool that is now integreated into the API Testing Component within the product APIConnect in Cloud Pak For Integration (CP4I) \href{https://www.ibm.com/common/ssi/ShowDoc.wss?docURL=/common/ssi/rep_ca/4/897/ENUS222-174/index.html}{2022.2.1}. For more details look at \href{https://community.ibm.com/community/user/integration/blogs/swetha-sridharan1/2022/07/13/introducing-auto-test-assist}{this blog}.

Autotest Assist has shown a lot of value in testing certain types of APIs during the development phase, before APIs are promoted to higher environments, uncovering several unique bugs in the areas of syntactic errors, missing implementation and also bugs that occur due to race conditions in the code. Once Autotest Assist is set-up, it requires minimal human intervention. Any subsequent effort is expended in triaging and debugging the failures reported by Autotest Assist.

Typically a test generation tool generates unique test cases that are then managed, scheduled and run manually by a testing framework. Autotest Assist is not a typical test generation tool. Instead of generating unique tests that then have to be run against the APIs, Autotest Assist  executes or exercises a sequence of tests, sending requests to the endpoint where the API is hosted. It can be seen an "exerciser" rather than as a test generation tool. Its output is an unbounded stream of API requests, which are generated one after another, in an online manner - each request depends on the internal state that Autotest Assist maintains, which is a reflection of Autotest Assist's understanding of the union of the requests and responses it has generated so far. Therefore, Autotest Assist does not generate (by default) unique, isolated test-cases, but rather a sequence of requests which exercises the System-Under-Test (SUT). The sequence of request generation will cease under two scenarios: a) a timeout has occurred (test result is "passed"), or b) an error was detected (test result is "failed").

Autotest Assist is a black-box tool, that does not tie itself to particular implementations or assumptions of APIs. That said, the tool can be adjusted to improve its coverage and bug-finding capabilities, making it more of a "grey-box" for certain APIs. This is done for APIs which justify such effort, for example, APIs that require a very high reliability. See section \ref{section:Customizations} on customizations.

To setup Autotest Assist for usage on a given API, Autotest Assist first takes the OpenAPI specification, and derives a specific model out of it. The model generated contains the resources which appear in the API, as well as the relations between them. We call this the "semantic model". Each resource is modeled, and is used as a basic building block for test generation. In addition, dependencies between resources are modeled: does resource A require another resource B to be created (making A dependent on B). For example, we may have 2 resources, \textit{Book} and \textit{Author}, such that creation of a \textit{Book} requires an \textit{Author} to be provided as an input, making the Book resource dependent on the Author. The semantic model is inferred using certain heuristics on the OpenAPI specification structure. Expanding the semantics of OpenAPI is a  \href{https://www.researchgate.net/publication/316653033_From_Open_API_to_Semantic_Specifications_and_Code_Adapters}{hot topic}, and we believe that Autotest Assist's generation of a semantic model alone may have a significant value for many users, not just in the API testing domain.

Autotest Assist has 2 modes of operation, sequential and concurrent. In the sequential mode, each request is generated only after the response of the previous generated request was received and checked. In this case, there is at most a single request in flight at all times across the system. Such a mode of execution maybe easier to debug, however this mode may hide underlying bugs related to atomicity, as concurrency is not exercised. On the other hand, in the concurrent mode, this restriction is removed, and several requests are executed in parallel in the system. This is of course a more complicated mode to generate requests and also to debug bugs. Its main advantage though is the ability to find non-atomic issues eg. race condition bugs, which manifest as a result of several requests accessing the same resource(s) in parallel.

Autotest Assist checking methods are:
\begin{enumerate}
    \item Syntactic checking of response body
    \item Check for HTTP Status Code 500 (i.e 5XX): "Internal Server Error"
    \item Semantic checking of responses - making sure the response content accurately matches the Autotest Assist's internal state. When dealing with asynchronous APIs, Autotest Assist needs to maintain partial internal states of API requests and responses to perform semantic checking. This feature of the tool is  work-in-progress and is not available as part of CP4I release \href{https://www.ibm.com/common/ssi/ShowDoc.wss?docURL=/common/ssi/rep_ca/4/897/ENUS222-174/index.html}{2022.2.1}
\end{enumerate}

In general, we see two potential classes of users of Autotest Assist: One class of users (typically Enterprise level customers) who dive deep into testing a specific set of APIs, adding manual modeling and customization in order to cover the entire depth of the API functionality using Autotest Assist. The second class of users who benefit more on quantity over quality, or breadth over depth when it comes to API testing: being able to test as many APIs and operations as possible, with minimal manual effort from their testing teams. Autotest Assist is developed with both types of users in mind. It allows for naive users to use the auto generated semantic model with minimal to no changes, at the same time for advanced users, the model can be customized to suit more specific requirements. 

Autotest Assist technology can be integrated into an existing testing framework by extending the framework to configure and invoke Autotest Assist. Typically in a test framework, a testing tool exists where Quality Assurance engineers write test cases to test specific APIs, the framework validates response with a known response and reports success/failures on each test case which are in turn grouped to a test suite.  These test suites can be scheduled to run at specific times during various stages in the development and release life-cycle. In such a framework, Autotest Assist tool can be integrated as a "test suite" that can potentially augment or in a more mature model, replace manual addition and updates of the test cases. Since Autotest Assist allows for customization of the areas to be tested, QA engineers can schedule Autotest Assist to generate requests on specific aspects of the API which may not have been covered by the manual tests initially. Additional aspects of "coverage" based testing is discussed in the conclusion section of this paper.

\subsection{Types of bugs uncovered by Autotest Assist}
In this section, we will describe a few bugs found by Autotest Assist in the sample \textit{Bookshop} application APIs as well as in the \textit{OpenShift} APIs in the order that they were found.  This will give an idea of the types of issues that can be uncovered by Autotest Assist. Although the examples are interesting they are anecdotal and do not represent all types of bugs that can be found by Autotest Assist.  

\subsubsection{API Schema mismatch errors}
Autotest Assist first found API Schema mismatch errors. When Autotest Assist was first run against the sample \textit{bookshop} application, Autotest Assist ended with a \textit{failed run}.  This was because a response received did not match the schema in the specification.  When Autotest Assist was first run on Openshift APIs, it got \emph{null} for the \emph{metadata.creationTimestamp} property, even though the schema did not allow it.

\subsubsection{Backend implementation errors}

Next, Autotest Assist exposed a few scenarios resulting in \emph{500} http error response, which indicates an "Internal Server Error", in the sample Bookshop application as follows:
Autotest Assist created a \emph{GET} request for non-existing customer and received \emph{500} http error response. Similarly, Autotest Assist received \emph{500} http error response when trying to delete a customer which was successfully created by a previous request. Additionally, Autotest Assist discovered an issue with the DELETE operation on the path "/api/v1/namespaces/{invalid name}/pods" of the Openshift APIs.  The issue is that the DELETE operation provided a good response even though Autotest Assist generated an invalid value for a parameter.

%syntactic bugs (upper/lower case etc.)
%random-nature bugs (1401-1450GB range not working correctly)
%500 bugs due to missing implementation
%500/syntactic bugs due to race condition
%Semantic checking bugs

%\section{Random test generation methodology}
%TBC it's s spectrum that deepens on the maturity of you automation.  The minimum is when you have automatic directed tests and the maximum is where all of your tests are random.
\pagebreak

\section{Methodology}

\begin{figure}[h]
\begin{center}
\includegraphics[width=0.9\textwidth]{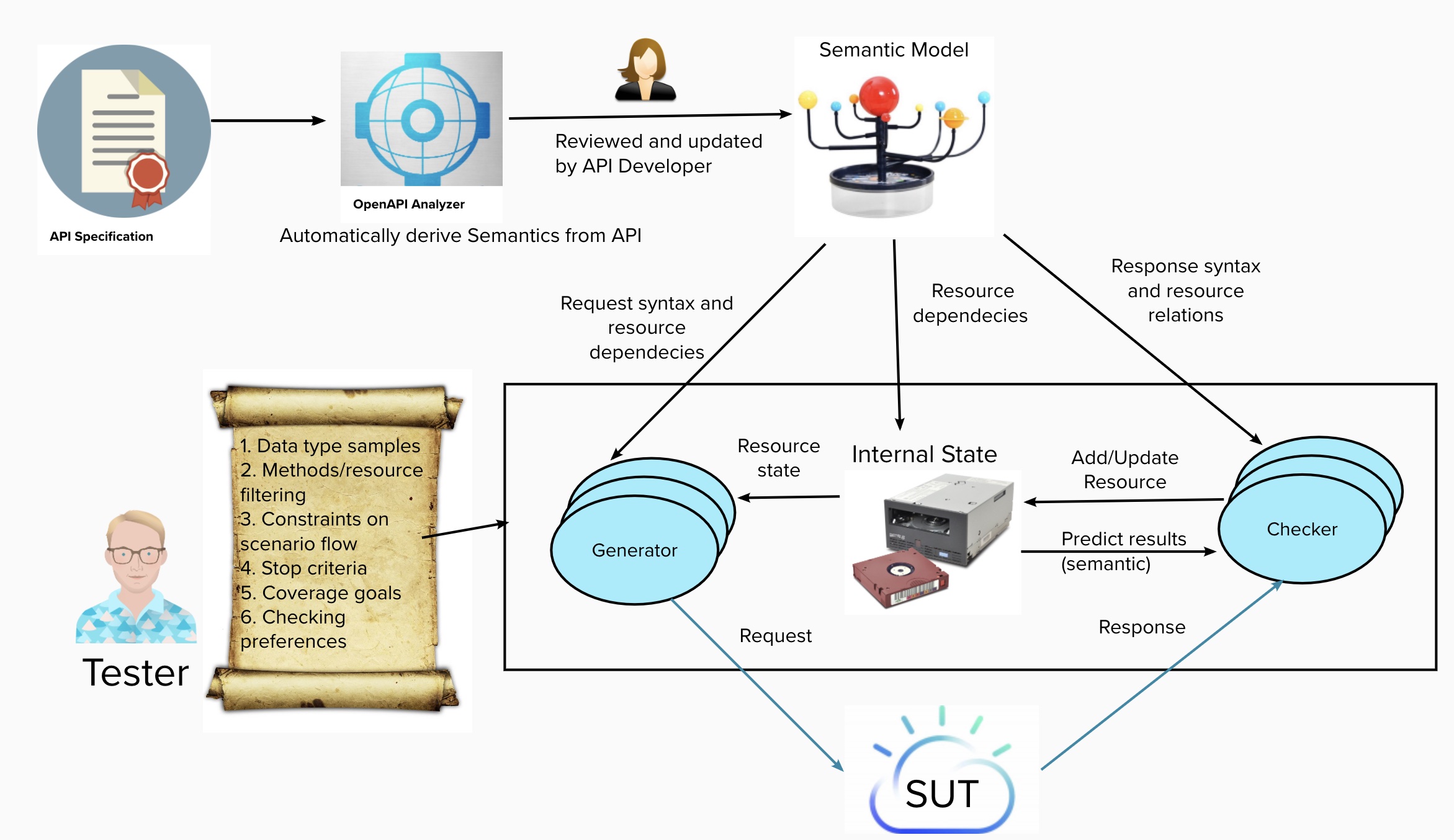}
\caption{\label{fig:Autotest Assist}Testing Methodology implemented in Autotest Assist}
\end{center}
\end{figure}

In the introduction section, we discussed the challenges of random test generation. 
In this section, we will elaborate on these challenges, and how they can be addressed using a testing framework, methodology and best practices outlined below. The testing methodology implemented in Autotest Assist is depicted in Figure \ref{fig:Autotest Assist}.  

\subsection{API Specification}
An API can define an endpoint URL where several resources can be accessed using a main API URL (root path) and multiple sub-paths. An Inter-API system is one in which each microservice defines a set of APIs for specific functions and a collection of microservices and their APIs together provide a complete application environment.  On the other hand, an Intra-API system is based on a set of operations and parameters that are used within the API itself. Systems can be either inter-API or intra-API or both. A single API endpoint can expose $n$ different operations on $m$ different resources. Typically, related functionalities and resources are grouped under one API which defines its boundaries and scope.

The correctness of the API requests require proper design and communication on both inter-API and intra-API systems. For example, the BookStore Application can expose the following APIs: Registration API, Locations API, Catalog API, OrderProcessing API and Payment API. Each of these APIs have interdependency between them, for example you cannot invoke the Payment API without an OrderId or you cannot access the OrderProcessing API before going through the Registration API and so on. In addition, each of these APIs in turn can have multiple resources and objects defined. For example, the Registration API can have two resources Customer and Book. Operations on these resources could be "CreateNewCustomer" "GetCustomerID", "GetBookID", "PostBookToCustomer", "DeleteBook" and so on. One cannot invoke the operation GetCustomerID without invoking CreateNewCustomer first. The order of invocation of the operations impact the underlying memory allocation strategies for the resources. These form the intra-API resources dependencies. As in the case of inter-API communication, the same set of challenges also exist for intra-API communication.  Again we need to ensure that dependencies on resources and operations are derived and understood well before a specific API can be exercised. The key to getting the resources and operations defined correctly in the API specification heavily relies on the design and documentation of the APIs.

Designing an API is more of an art. There is no one right or wrong way to do it. However, there are some basic rules and best practices that would help guide towards designing APIs that are readable, testable and extensible. \href{https://cloud.ibm.com/docs/api-handbook}{IBM Cloud API Handbook} is a good resource. The API specification should ensure to reflect the correct resource allocation model of the underlying resources. For example, the specification should define an API path to perform a "CREATE" operation on a new resource which allocates a unique ID to the resource. This ID should then be referenced in the API path to perform the "DELETE" operation. 

Correct resource allocation, usage and release is an important part of API's quality.  To enable the testing of problems related to resource allocation and usage, the API specification needs to define them.  There are ML technologies that tries to infer the dependencies between resources and their operations, but starting with a good API design with explicit definitions sets the foundation for a strong and effective API testing.
    
\subsection{Semantic Model}
A semantic model is derived based on the OpenAPI specification. A well designed API yields a closer to accurate model derived in an automated manner - we call this automation framework an OpenAPI Analyser. Autotest Assist has implemented such an analyser. The analyser can extract the resource dependencies from the APIs using the names and descriptions of the parameters specified. It can use Natural language processing (NLP) functions to determine if two or more operations are dependent on each other based on the parameters names they are referring to.  In such a case the output from one operation is fed as input to the other.
The analyser builds a file that can be reviewed and updated by the API Developer who is expected to have a better understanding of the dependencies of resources used by the API. The semantic model output also indicates to the API Developer how well the API has been designed. If the API design is not well done, this gives an opportunity to the developer to revise the APIs specification in accordance with the best practices. 
%A sample semantic model derived from an OpenAPI specification is shown in \ref{fig:sematnic}

\subsection{Requests Generator and Checker}
\label{subsec:gen_and_chk}
The Generator and Checker form the core components of the Autotest Assist framework. 
The semantic model provides a resource-centric view of an API, which indicates all the operations defined in an API attributed to a specific resource they operate on. For example, a bookshop API contains a resource called \emph{Book} and there are operations which allow to create a book, delete a book, modify or retrieve a list of existing books. With this semantic model as the input, the Generator component first randomly selects a resource from a list of resources defined in the model, then it randomly selects an operation from a list of resource operations. Once an operation is selected, the Generator has to generate values for operation parameters. Before Autotest Assist starts request generation, it analyzes an OpenAPI specification and creates a sampling specification for each operation. The sampling specification defines domain and value distribution for each operation parameter. When the Generator needs to specify values for operation parameters, it requests a single sample from the sample specification.

Once a response is received, Autotest Assist checks if its status code is defined in the specification. If the status code is not defined, the Checker reports an error. For example, all \emph{5XX} status codes are reported as errors by this checker. If the status code is defined in the specification, Autotest Assist validates the response against a schema defined in the specification for this status code and response content type. The above checks belong to syntax checking and performed directly against the specification. Additionally, Autotest Assist can perform semantic checking, where it tracks resource state and predicts what status code has to be reported. For example, if the bookshop application returns positive status code for a book which was removed, the semantic checker could report an error.

\subsection{Internal States}
Autotest Assist stores information regarding resources it creates or discovers in internal states. The information is updated not only by operations which report resource description, such as \emph{GET} or \emph{POST}, but also by operations which modify or delete a resource. 

The state information allows to get good responses for requests which has to provide resource ids in their input. For example, an operation which creates an Order in the bookshop application has to specify an id of a customer and ids of books which were ordered. To create such a request, the Generator requests ids of all customers in the internal state and randomly selects one of them. Additionally, the Generator also requests ids of the Book resources and randomly selects one or more ids to create a request. 

The internal state also allows Autotest Assist to perform semantic validation as described in \ref{subsec:gen_and_chk}.

\subsection{Customizations}
\label{section:Customizations}
Any random test generator framework is successful only when it exposes and allows the customization of inputs and configurations of the internal framework components. In this random testing methodology and framework, although the system randomly generates tests, it allows for the user of the framework (API Developer or tester) to specify which areas of the APIs the framework should put more focus on. For example, the API tester can customize the tool to add more weight to the PUT operation over the GET operation in which case PUT requests are generated proportionally more than the GET requests. Other areas of customziations can include error handling, duration of the tests (can range from 2 minutes to 20 hours), stop/exit criteria and so on. 
\subsection{Debugging the System Under Test (SUT)}
Regression testing is the process of running a set of existing functional and non functional tests in order to determine if a change to the System Under Test (SUT) due to a bug fix or software enhancement caused a problem. A regression test is automated if the results of each tests, i.e., whether it succeeded or failed, are determined programmatically.  In such a case, once the regression is run, a report is automatically produced that states if each of the tests that constitute the regression passed or failed.  An automatic regression on a SUT increases the confidence of the development team in their ability to fix and progress the SUT while keeping its design clean. This in turn increases the life span of the software and avoids accumulating technical debt.

It is a best practice to enhance an automatic regression with tests that test for bugs that were found and corrected in the software. This needs to be done with care to avoid accumulating essentially duplicate tests and ending up with a regression that takes too long to run.   

When random test is used it may be desirable to follow the same practice of adding individual tests generated by the random test generation framework into the regression test bucket. This process is however non trivial. The random generation process may have run for a long time until the problem emerged. Thus, simply re-running the random test generation sequence of API calls as part of the regression is not viable.  In addition, the random test generator does not produce a re-runnable set of operations. The good news is that in order to debug a sequence of API calls produced by the random test generation, the developer can create a smaller, minimal, execution that could produce the same problem. This debugging step is sometimes referred to as "recreate" and is many times necessary in order to debug the problem.  In order to enable the recreate process, the random test generation saves the sequence of API calls that lead to the failure.  The recreated test is the test that could be added to the regression.

At a higher maturity level one could trust the random generation process completely and drop the need for a regression test. This includes the development of some statistics of how long the random test generation needs to run in order to reveal a bug of the sort that was revealed by the random generation before the fix.  Intuitively, the random test generation needs to run long enough so that it is unlikely that the recreated test will not occur. For example, if the random test generation is calling 10 APIs, and the bug is revealed by calling one of them, then we need to randomly generate tests so that it is unlikely that an API  will not be invoked. In other words, we will need to run the random test generation, and pick up an API to execute at random, on the order of a 100 or more times.

\section{Conclusion}
In this paper we discussed Random Test Generation which introduces a different perspective and approach on testing API endpoints than the traditional test automation frameworks. 
The random test generator methodology uses the Open API specification to derive a semantic model of the resources used within an API, generates test requests, maintains internal state of requests and responses received, uses the internal state and a sampling space to generate newer requests continuously and intelligently deduce syntactic and semantic errors of the APIs as well as the backend systems. 

The random test generation fits perfectly in an API development phase, though nothing prevents it from using it in a production phase as well. The framework can be extended to track the coverage of the APIs exercised. By providing a view of the number of the APIs requests, operation and parameter values, the framework can give a powerful insight into the areas that lack coverage. By classifying the areas of desired coverage based on a domain, for example, security or OWASP top 10 security venerability, performance related test cases, race conditions related test cases, the model can become highly extensible and can be focused on specific areas. We expect this methodology to fit well within an existing testing dashboard alongside any existing test automation framework. In an established and mature implementation, it has the potential to replace the traditional management of test cases entirely.

\bibliographystyle{alpha}
\bibliography{main}

\end{document}